\documentclass{aa}
\def\et{et al.}
\def\deg{$^\circ$}

\def\aap{A\&A}

\def\aj{AJ}

\def\sci{Sci}

\def\rsci{Radio Science}
\def\jgr{J.\ Geophys.\ Res.}

\begin{document}

   \thesaurus{11         
              (02.16.1;  
               03.01.2;  
               03.13.5;  
               03.20.2;  
               05.01.1)}  
\title{A GPS-based method to model the plasma effects in VLBI observations}

   \author{E.\ Ros\inst{1,2}
\and
J.M.\ Marcaide\inst{1}
\and
J.C.\ Guirado\inst{1}
\and
E.\ Sard\'on\thanks{%
{\it Present address:\/} Grupo de Mec\'anica del Vuelo S.A.,
Isaac Newton 11, PTM Tres Cantos, E-27860
Madrid, Spain}\inst{3}
\and
I.I.\ Shapiro\inst{4}
          }

   \offprints{E.\ Ros, ros@mpifr-bonn.mpg.de}

   \institute{
Departament d{'}Astronomia i Astrof\'{\i}sica,
Universitat de Val\`encia, E-46100 Burjassot, Val\`encia, Spain
\and
Max-Planck-Institut f\"ur Radioastronomie,
Auf dem H\"ugel 69, D-53121
Bonn, Germany
\and
Deutsche Forschungsanstalt f\"ur Luft- und Raumfahrt e.V., Fernerkundungstation
Neustrelitz, D-17235 Neustrelitz, Germany
\and
Harvard-Smithsonian Center for Astrophysics, 60 Garden St.,
Cambridge, MA 02138, US
          }

\date{Received 17 September 1999 / Accepted 17 January 2000} 
\titlerunning{GPS-method to model plasma effects in VLBI observations}

\maketitle

\begin{abstract}
Global Positioning System (GPS) satellites broadcast at frequencies
of 1,575.42\,MHz (L1) and of 1,227.60\,MHz (L2).  The dispersive property of 
the ionosphere can be used to combine independent measurements
at the two frequencies to estimate
the total electron content (TEC) between a GPS receiver
site and a broadcasting satellite.  Such measurements, made at
sites near to Very Long Baseline Interferometry (VLBI)
sites, can be used to estimate the
ionospheric contribution to VLBI observables.
For our 1991.9 astrometric VLBI experiment in which we obtained 
group-delay observations in the 8.4 and 2.3\,GHz bands simultaneously, we
found that the GPS and VLBI determinations of the ionosphere delays 
agreed with root-mean-square differences below 
0.15\,ns for intercontinental baselines and 0.10\,ns for continental
ones.  We also 
successfully applied the GPS-based procedure
to reduce the ionospheric effect in phase delays used for high 
precision differenced astrometry at 8.4\,GHz for this same experiment.
\end{abstract}


\section{Introduction}

VLBI provides unprecedentedly accurate angular resolution 
through observations of celestial bodies with 
radio telescopes spread over the Earth's surface.  
Each observing
station records data on magnetic tapes.  The local-oscillator
signals and the time-tagging of the data are governed by hydrogen-maser
frequency standards.  The tapes are processed
in special-purpose correlators to determine the so-called VLBI observables: 
group and phase delays and phase-delay rates.

A main problem in determining the sky positions of celestial radio sources 
from these VLBI observables is the effect of the Earth's ionosphere on them.
The use of GPS satellite data to remove this effect forms the thrust of
this paper.

The ionosphere is characterized by
its content of free electrons and ions. 
The F$_2$ layer of the ionosphere has the largest density of charged particles,
with values up to $3\times10^{12}$\,m$^{-3}$.
The total electron content per square meter along a line of sight
is the number of electrons in a column of
one square meter cross section along the ray path:  
\begin{equation}
{\rm TEC}= \int_0^{h_0} N\cdot dh,
\label{equ:tec-def}
\end{equation}
where $N$ is the spatial density of electrons,
$h$ is the coordinate of propagation of the wave, and $h_0$ corresponds
to the effective end of the ionosphere.
TEC is highly variable and depends on several factors,
such as local time, geographical location, season, and solar activity.
TEC can have values between 1 TECU (or TEC unit, defined as
10$^{16}$\,m$^{-2}$) and 10$^3$ TECU.  Epochs of greater solar
activity cause higher values of the TEC.

The ionosphere affects the phase and group delays oppositely (to first
order, see, e.g.,
Thompson \et\ \cite{tho86}):
\begin{equation}
\Delta \tau = \mp \frac{\kappa}{c\nu^2}\cdot {\rm TEC},
\label{equ:ioeffe-1}
\end{equation}
where $\kappa\approx40.3$\,m$^3$s$^{-2}$, 
$c$ is the speed of light (m$\cdot$s$^{-1}$), 
and $\nu$ the frequency (Hz), and where we neglect magnetic field effects 
and assume $\nu$ is large compared with the local plasma frequency
(for an extreme case, the plasma frequency is of $\sim$15\,MHz). 
The negative sign applies for phase delays and the positive sign for 
group delays.
In standard astrometric VLBI experiments observations are made simultaneously
in two well separated bands of frequencies in order to
estimate the ionospheric effect.  A nearly vacuum equivalent delay
can be obtained from the following expression:
\begin{equation}
\tau_{\rm free}=\frac{(\nu_1/\nu_2)^2\tau_1 - \tau_2}{(\nu_1/\nu_2)^2-1},
\label{equ:iofree-delay}
\end{equation}
where $\tau_i$ is the delay --either group or phase-- 
at frequency $\nu_i$ ($i=1,2$, $\nu_1>\nu_2$). 

Thus, with dual frequency observations, the ionosphere effect 
can largely be removed from the VLBI data.  Such 
removal can also be made for single-frequency observations
via Eq.\ (\ref{equ:ioeffe-1}), if 
estimates of the TEC along the lines of sight of the radio telescopes 
are available from other observations.

Guirado \et\ (\cite{gui95}) showed that it is possible to estimate
the ionospheric effect with accuracy useful for astrometric purposes from 
Faraday-rotation measurements.
In this work, the authors used a ``clipped" sinusoidal function to
model the diurnal behavior of the TEC. 
In this model, the night component is 
constant and equal to the minimum TEC value, and the day component is
expressed as the positive part of a sinusoid, with its 
maximum some hours after noon.
Their observations were obtained in late 1985, a
time of minimum solar activity.

In the method presented here we used 
GPS measurements that provide 
TEC values as a function of time.
Such GPS-based TEC determinations were first successfully applied to geodetic
VLBI by Sard\'on \et\ (\cite{sar92}).

\section{The Global Positioning System and the TEC.\label{sec:gps-system}}

An introduction to the Global Positioning System (GPS)
can be found in Hofmann-Wellenhof \et\ (\cite{hof97}).
A main use of the GPS system is
to determine the position
$(x,y,z,t)$ of a GPS receiver on Earth's surface.
The system consists of a constellation of 24 satellites broadcasting
electromagnetic signals in two narrow frequency bands, 
a set of monitoring ground sites, the Master Control Station,
and GPS receivers. The 24 satellites orbit the Earth in near
circular orbits with a 12\,hr period, at a height of about 20,200\,km, 
and an inclination of 55\deg.  The spacecraft are in six
orbital planes with four satellites nearly equally spaced along the orbit 
in each plane.
At any moment, from any point on Earth, it is possible to detect
signals simultaneously from 7 to 9 of these satellites.  
Each satellite broadcasts a block of data
every 30 seconds, consisting of a description of its orbit and of
GPS time, as well as a pseudo-random code every
millisecond (coarse-acquisition C/A, for civilian
use) and a precision code (P, 266\,d period, for military use), 
usable to determine more accurately the position of the ground receiver.

The oscillators of the satellites generate a fundamental frequency
$\nu_0=10.23$\,MHz, which is the P-code frequency.  The C/A-code frequency
is $\nu_0/10$.  The GPS signal is emitted at two frequencies,
154\,$\nu_0$ and 120\,$\nu_0$ (L1 and L2, respectively, 1,575.42\,MHz and
1,227.60\,MHz, or $\lambda\lambda$19 and 24.4\,cm).
L2 carries only the P signal,
and L1 both P and C/A signals.

\subsection{The GPS observables.\label{sec:gps-obs}}

The main GPS observable is
$dT$, the time of transit of the signal from the satellite to
the ground receiver.
The value of $dT$ can be determined, in effect, 
by comparing the time of broadcasting of
the codes P or C/A with the time of receipt of these codes.
An observable that corresponds to $dT$ 
can be obtained by cross correlation of the signal 
received from the satellite at each band with a reference signal
generated at the receiver and ``tied" to GPS time.   
In this case, $dT=\phi/2\pi\nu$, where $\phi$ is the total phase change
of the signal during propagation from satellite to receiver and $\nu$ is
the center frequency of this transmitted signal.
This observable is the 
phase delay of the carrier signal.

The GPS observable is affected by the following: the ionosphere,
the troposphere, $2\pi$ phase ambiguities (equivalent to multiples of
of 634.75\,ps and 814.60\,ps, respectively, for L1 and L2),
multipath (e.g., from signals that
are reflected or scattered into the receiver antennas from objects 
nearby to it), different effective location of
receivers for L1 and L2, instrumental delays (different for L1 and L2), 
degraded coding of the signals, and clock errors.
The contributions to the observables of the largest of these effects 
can be sharply reduced by application of 
suitable techniques (see, e.g., Blewitt \cite{ble90}).

\subsection{Obtaining the TEC from GPS data.\label{sec:gps-equ}}

The GPS observable
can be modeled as a function of distance from satellite to receiver,
ionospheric delays, tropospheric delays, clock 
errors, and instrumental phase- and group-delay biases
(Sard\'on \et\ \cite{sar94}).
The ionospheric term can be estimated by a combination of
the L1 and L2 observables.
We can denote the TEC for any observation direction as $I^i_k(t)$.  This TEC 
is defined along the line of sight from
the radio telescope $k$ 
to the radio source $i$ and can be expressed approximately 
as a function $\cal V$ (vertical
value of TEC) of time $t$ and the intersection point $P^i_k$ (``ionospheric
point") of the line of 
sight from $k$ to $i$ with the (average) 
F$_2$ layer of the ionosphere 
(at an altitude of $h_{{\rm F}_2}=350$\,km), times
the obliquity or slant function $S(e^i_k)$, defined as the secant of the zenith
angle at the ionospheric point (see below
for geometry clarification), which is a function of the 
elevation angle $e^i_k$ of the observation:
\begin{equation}
I^i_k(t)=S(e^i_k)\cdot {\cal V}(P^i_k,t).
\label{equ:tec-form}
\end{equation}

The positions of the involved sites and the ionospheric points
can be expressed in a 
geocentric coordinate system 
$(X,Y,Z)$ or in a geocentric-solar one $(\Psi,\chi,Z^\prime)$ with the 
$Z^\prime$ axis directed toward the Sun from the Earth center, 
$\Psi$ the angle
in the $XY$-plane (measured counterclockwise from  $X$) and $\chi$ the 
angle with apex at the Earth's center, measured
from the direction to the Sun ($Z^\prime$) to the direction to 
$P^i_j$.
This latter coordinate system is useful 
since the ionosphere is roughly time-independent
in this reference frame.
Following Sard\'on \et\ (\cite{sar94}), we replace $\cal V$ 
for a GPS site $j$ and satellites $l$ by its locally
linear approximation in $P^l_j$ (ionospheric
point towards satellite $l$) using the $(\Psi,\chi,Z^\prime)$-coordinate 
system: 
\begin{equation}
{I}^l_j(t)=S(e^l_j)\cdot [A_j(t)+B_j(t)d\Psi^l_j(t) +
C_j(t)d\chi^l_j] + K^l + K_j.
\label{equ:tec-terms}
\end{equation}
Here, the coefficients for each site are
$A$ (0$^{\rm th}$ order, vertical), $B$ 
(1$^{\rm st}$
order, $\Psi$-direction), $C$ (1$^{\rm st}$ order, 
$\chi$-direction).  We have also introduced the 
instrumental
GPS satellite $K^l$ and receiver $K_j$ biases.
$A_j(t)$ is the
vertical TEC at site $j$, $d\Psi^l_j(t)=\Psi^l_j-\Psi_j$, and 
$d\chi^l_j=\chi^l_j-\chi_j$ are the coordinates
$\Psi$ and $\chi$ of the ionospheric point 
of the line of sight from $j$ towards the satellite $l$ minus 
the corresponding coordinates
of the GPS site $j$.
The coefficients $A$, $B$, $C$, can be determined, and the $K$-biases
largely removed, using a Kalman filtering method (Herring \et\ \cite{her90})
with the data from different satellites $l$ (about 8 at a time) to obtain 
an estimate of the TEC from GPS data.  The $K$ biases 
can be due to such effects as errors in the estimation of 
phase ambiguities and multipath (see Sard\'on \et\ \cite{sar94}). 
In sum, we produce estimates of the total electron content
of the ionosphere using data from a network of GPS stations
and a multiplicity of satellites.

\section{Vertical TEC from GPS and TEC for VLBI.\label{sec:gps2vlbi}}

Here we describe the formulas we used to estimate the TEC 
along the paths of a VLBI observation from the values of the vertical
TEC (evaluated by the method described in Sect.\ \ref{sec:gps-equ}) 
at a GPS site near each one of the VLBI sites.

\begin{figure*}[htb]
\vspace*{308pt}
\includegraphics{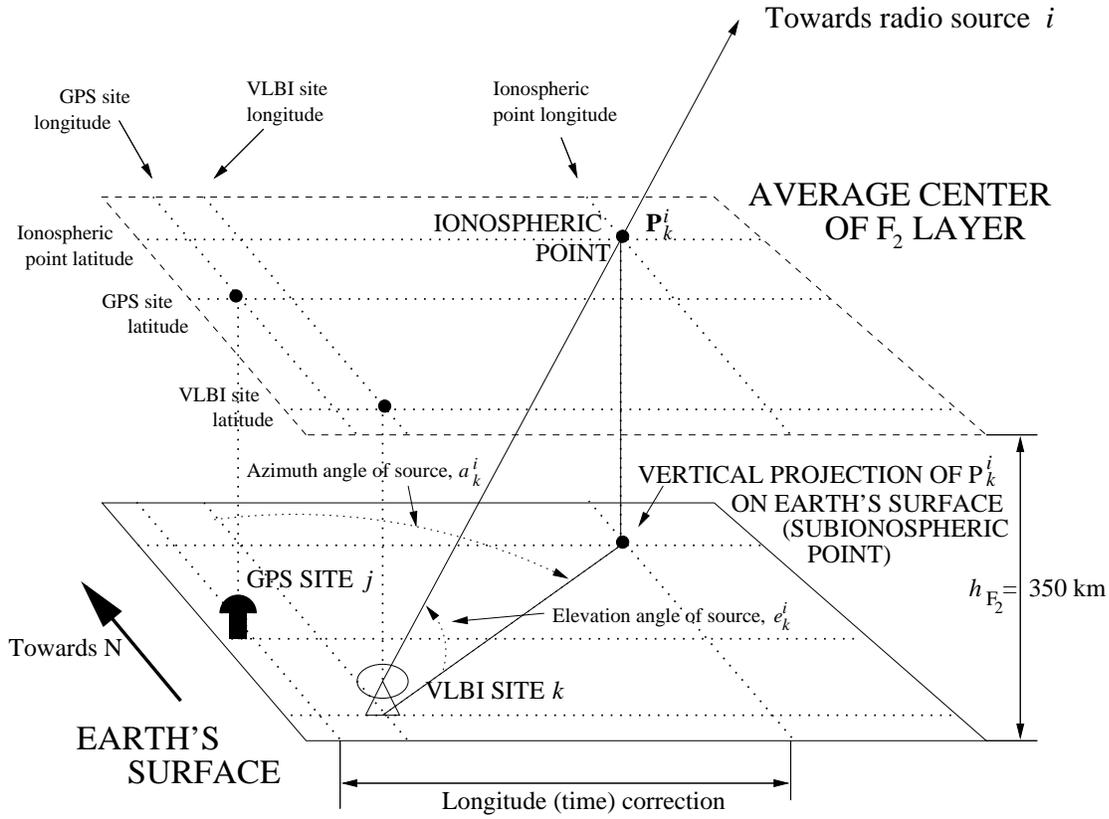}
\caption{Geometry used to estimate TEC from GPS measurements (see text).
The vertical lines here are radial (from the Earth's 
center) and the tangent planes drawn
are approximations to the surface of spheres (Earth and F$_2$ layer of
ionosphere) having, e.g., the VLBI site and its ionospheric
point as points of tangency.
\label{fig:geom-iono}
}
\end{figure*}

\begin{figure}[htbp]
\vspace{135pt}
\includegraphics{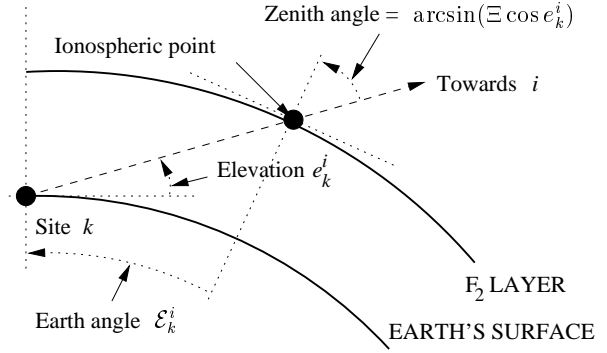}
\caption{Geometry of the obliquity or slant factor, which is the secant of
the zenith angle, defined by the elevation $e^i_k$ (see text for definition
of $\Xi$).  Curvatures
of the Earth's surface and the ionosphere 
F$_2$ layer are exaggerated for clarity.
\label{fig:slant}
}
\end{figure}

Consider a GPS site $j$ at latitude $\zeta_j$ and 
longitude $\lambda_j$, and a VLBI site $k$ at position 
$(\zeta_{k},\lambda_{k})$ (see Fig.\ \ref{fig:geom-iono}).
The coordinates of $P^i_k$, denoted by $(\zeta_{P^i_{k}},
\lambda_{P^i_{k}})$
at the epoch the radio source is at elevation $e^i_k$ and azimuth $a^i_k$,
are
(Klobuchar \cite{klo87}):
\begin{eqnarray}
\zeta_{P^i_{k}} & = & 
\arcsin ( \sin \zeta_{k} \cos {\cal E}^i_j + 
          \cos \zeta_{k} \sin {\cal E}^i_j \cos a^i_j)\\
\lambda_{P^i_{k}} & = & 
\lambda_{k} + 
\arcsin \left(\frac{\sin {\cal E}^i_j \sin a^i_j}{\cos \zeta_{P^i_{k}}}\right), 
\label{eq:lambdak}
\end{eqnarray}
where $(-\pi/2 \le \zeta_k \le \pi/2)$ and
the $\arcsin$-function in $\lambda$ holds for values in
the interval $(-\pi/2,\pi/2)$, appropriate for Eq.\ (\ref{eq:lambdak})
since GPS and VLBI sites are nearly collocated.
${\cal E}^i_k$ is the angle, measured from the center
of the Earth between the line to site $k$ and the line to the 
ionospheric point for radio source $i$:
${\cal E}^i_k=\pi/2-e^i_k-\arcsin(\Xi \cos e^i_k)$ (see
Fig.\ \ref{fig:slant}, where 
$\Xi=R_\oplus/(R_\oplus+h_{{\rm F}_2})$, $R_\oplus$ is the Earth's
radius, and $\Xi\approx 0.948$ for $h_{{\rm F}_2}=350$\,km).
The local time at the 
ionospheric point is $t=(\lambda_{P^i_{k}}/15)+$UT\,hr ($\lambda$ in
degrees, with $\lambda$ positive to the East).

We assume
that the vertical TEC $A_{P^i_{k}}(t)$ at the ionospheric point is
given in terms of $A_j(t)$ by:
\begin{equation}
A_{P^i_{k}}(t)=A_j\left(t+\frac{\lambda_{P^i_{k}} - \lambda_j}{15}\right).
\end{equation}
This relation effects a longitude correction but ignores any latitude dependence
of the TEC values.
This approach is  reasonable for mid-latitude stations
and sources at high declination since GPS sites can be
collocated with VLBI sites, or at least placed relatively near to them, 
and since TEC changes more rapidly with longitude than with latitude.

The slant factor was defined above as the secant of the zenith
angle, which is $\arcsin (\Xi \cos e^i_k)$ (see Fig.\
\ref{fig:slant}, and Klobuchar \cite{klo87}).
Thus, the TEC at the ionospheric point $P^i_{k}$, mapped by 
the slant factor (Eq.\ \ref{equ:tec-form}) gives
\begin{equation}
I^i_{k} (t)= \sec\{\arcsin (\Xi \cos e^i_k)\} \cdot 
A_{P^i_{k}}(t),
\end{equation}
which, with Eq.\ (\ref{equ:ioeffe-1}) yields our estimate
of the ionospheric delays at the indicated VLBI site.  The overall ionospheric
effect on a VLBI observable is a simple linear combination of the effects from 
each of the two sites involved in the observation.

\section{A case to test the method: VLBI observations
of the Draco triangle.}

Progress in high precision phase-delay difference
astrometry has been made by Ros \et\ (\cite{ros99}) through VLBI
observations of triangle formed by the radio sources
\object{BL\,1803+784}, \object{QSO\,1928+738} and \object{BL\,2007+777}, 
in the Northern
constellation of Draco (the Dragon).  The observations were made
simultaneously at the frequencies of 2.3 and 8.4\,GHz
at epoch 1991.89 with an intercontinental interferometric array.
The angular separations among these radio sources were determined
with submilliarcsecond accuracy from a weighted-least-squares 
analysis of the differenced and undifferenced phase delays.
The modeling of these 
astrometric VLBI observations was sufficiently accurate to 
estimate reliably the ``$2\pi$ ambiguities" in the differenced phase delays for 
source separations of
almost 7$^\circ$ on the sky.  For such angular distances,
this accurate ``phase connection"
at 8.4\,GHz, yielding the phase delays with standard errors
well within one
phase cycle over the entire session of observations,
was demonstrated
at an epoch of solar maximum. 
As in earlier works (e.g., Guirado \et\ \cite{gui95}, Lara \et\ 
\cite{lar96}), after phase connection 
the effects of the extended
structure of the radio sources were largely removed from the phase-delay
observables.  The effects
of the ionosphere
were also mostly removed, via the GPS-based method described in this paper.

\begin{figure}[htb]
\vspace*{225pt}
\includegraphics{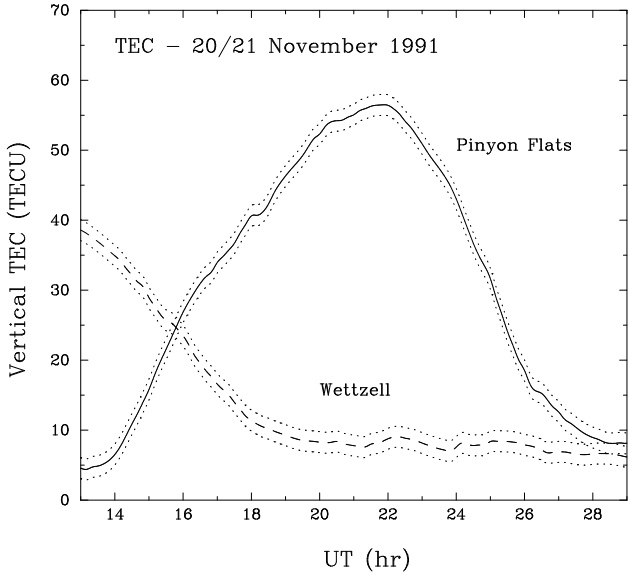}
\caption{Values of the vertical
TEC at the GPS sites of Pinyon Flats (CA, US) and
Wettzell (Germany)
(Ros et al.\ 1999).  These values were used to compute the TECs
at nearby VLBI sites.  
We assumed an error
of 1.5\,TECU (represented by the dotted lines), much 
larger than the statistical standard errors.
\label{fig:gm6-tec}}
\end{figure}

In 1991 the number of available GPS sites was small, and only 
data from Goldstone and Pinyon Flats in the US, and from
Herstmonceux and Wettzell in Europe (see Table \ref{tab:gps-vlbi} for 
details) were available to be used for our experiment.
GPS data from the two US sites were
used for the VLBI sites at 
Fort Davis (TX), Pie Town (NM), Kitt Peak
(AZ), and Los Alamos (NM) to estimate the TEC for observations at these 
sites.
Similarly, GPS data from both European sites were used for
Effelsberg (Germany).
The VLBI observations were carried out from 14\,hr\,UT 
on 20 November to 4\,hr\,UT on 21 November 1991.  In Fig.\  
\ref{fig:gm6-tec} we
show the corresponding vertical TEC values for one of the 
GPS sites on each continent.
We see the dusk and night part of the data for Wettzell, and
the daylight data for Pinyon Flats.

\begin{table}[htb]
\caption[]{Positions of the GPS sites and the VLBI sites (Ros \et\ \cite{ros99})
and maximum and minimum values of the vertical
TEC value on 20/21 November 1991 (see text for definitions of 
notation).\label{tab:gps-vlbi}}
\begin{tabular}{@{}l@{~}r@{~}r@{~}c@{~~}c@{~}l@{~}r@{~}r@{}}\hline
GPS             & $\zeta_j$ & $\lambda_j$ 
                                       & $A_j^{\rm min}$ 
                                            & $A_j^{\rm max}$  
                                                  & VLBI       & $\zeta_{k}$    
                                                                         & $\lambda_{k}$  \\ 
site            &          &          & {\tiny [TECU]} 
                                            & {\tiny [TECU]}  
                                                  & site    &         &          \\ \hline
{\tt DS10}$^1$  & 35.2\,N  & 116.9\,W & 5  & 57  & {\tt FD}$^a$& 30.6\,N & 103.9\,W \\
{\tt PIN1}$^2$  & 33.6\,N  & 116.5\,W & 5  & 57  & {\tt PT}$^b$& 34.3\,N & 108.1\,W \\
                &          &          &    &     & {\tt KP}$^c$& 32.0\,N & 111.6\,W \\
                &          &          &    &     & {\tt LA}$^d$& 35.8\,N & 106.2\,W \\ \hline
{\tt HERS}$^3$  & 50.9\,N  &  0.3\,E  & 8  & 38  & {\tt EB}$^e$& 50.3\,N &   6.8\,E \\
{\tt WTZ1}$^4$  & 49.1\,N  & 12.8\,E  & 8  & 39  &             &         &         \\ \hline
\end{tabular}
\noindent
\\
{\footnotesize
$^1$~Goldstone Deep Space Tracking Station, NASA, CA, US;
$^2$~Pinyon Flats Observatory, SIO, UCSD, CA, US;
$^3$~Herstmonceux, East Sussex, England;
$^4$~K\"otzting, IFAG, BKG, Bavaria, Germany.\\
$^a$~Fort Davis, VLBA, NRAO, TX, US;
$^b$~Pie Town, VLBA, NRAO, NM, US;
$^c$~Kitt Peak, VLBA, NRAO, AZ, US;
$^d$~Los Alamos, VLBA, NRAO, NM, US;
$^e$~Effelsberg, MPIfR, North-Rhine-Westphalia, Germany.
}
\end{table}

From the GPS-based estimates of the TEC along the lines of sight
for each VLBI site, we 
calculated the ionospheric contribution for each baseline and epoch
of observation.
These ionospheric contributions were removed from the VLBI observables.
For the phase delays, these contributions 
ranged in magnitude from 0 to 1.2\,ns
at 8.4\,GHz for intercontinental baselines, and were less than 0.1\,ns for
continental baselines.  The intercontinental baseline lengths range 
from $\sim$7800 to
$\sim$8300\,km, and the differences in local 
time are $\sim$7.5\,hr
(see Fig.\ \ref{fig:gm6-tec}).  Since the ionospheric
effect is the combination of effects for both antennas, the ionospheric
delay is quite important
for our intercontinental baselines.
By contrast our US-continental baseline lengths
range from $\sim$200 to $\sim$750\,km which corresponds at most to
about 30 minutes
difference in local time and a much smaller ionospheric effect on
the VLBI observables.

\begin{figure}[htb]
\vspace*{260pt}
\includegraphics{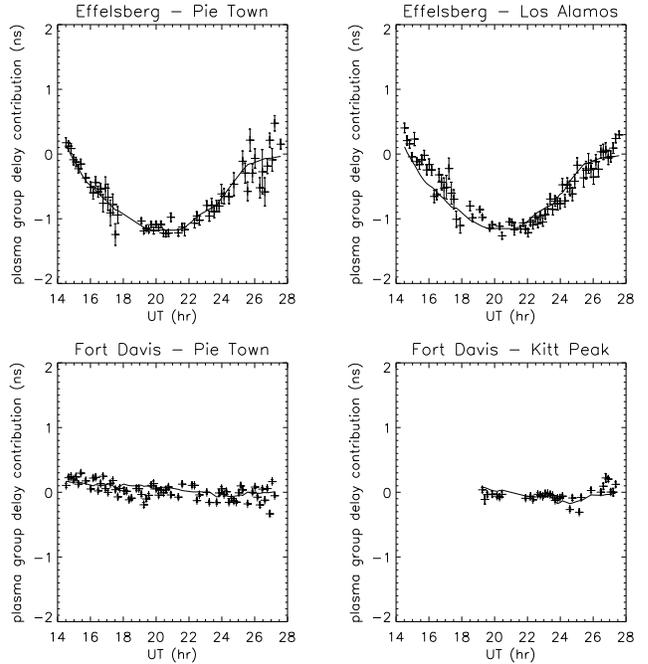}
\caption{Comparison between the ionospheric delays at 8.4\,GHz
estimated by the combination of group delays at 8.4 and 2.3\,GHz from 
the VLBI observations of Ros et al. (\cite{ros99}) (points with error bars) 
and the corresponding ionospheric delays
estimated by the GPS TEC determination reported here (solid line).  
Data from four representative baselines of a total of ten from the 
VLBI experiment are shown here.  Root-mean-square values for the
differences are, respectively, of 0.11, 0.15, for the upper panels 
and of 0.10, 0.08\,ns for the lower panels.
\label{fig:gm6-gps-gde}}
\end{figure}

Eq.\ (\ref{equ:iofree-delay}) provides the usual way to largely
remove the effect of the ionosphere
on dual-band VLBI observations. 
The phase delays from our 2.3\,GHz observations
could not be freed from $2\pi$ ambiguities due to the large
scatter in these data.
However, unambiguous but
less precise group delays were available at both 2.3 and
8.4\,GHz.  The ionospheric contributions were estimated from
these group-delay data.  In Fig.\ \ref{fig:gm6-gps-gde} we 
show the comparison 
of GPS-based and dual-band VLBI-based ionospheric delay estimates
at 8.4\,GHz.
We show four of the ten available baselines as representative examples: 
two intercontinental and two continental-US ones.
Apart from the larger dispersion in the
group-delay than in the GPS-based data, 
this comparison provides a good independent confirmation of
the reliability of the GPS-based method for the correction of VLBI data.  
The error bars for the group delays shown in the 
figure are the appropriate combination of the 
statistical standard errors for the data at these frequencies 
(see Eq.\ \ref{equ:iofree-delay}).
The statistical standard errors for the GPS 
estimates are each about 0.2\,TECU.  We
assume a much larger standard error --1.5\,TECU-- to try to account
for possible inaccuracies not estimated with the Kalman filtering, such
as incorrect values for $h_{{\rm F}_2}$, the consequent changes in
the mapping function and in the eventual position of the ionospheric
point, and other unmodeled effects.  Thus, we infer
a corresponding contribution of $\sim$30\,ps to the standard errors of the 
phase delays at 8.4\,GHz.
For the data presented in Fig.\ \ref{fig:gm6-gps-gde}, 
the root-mean-square
of the differences between 
the results from the two
methods are below 0.15\,ns for intercontinental baselines and 
0.10\,ns for continental ones.

\section{Conclusions}

Assuming that the ionosphere can be modeled usefully as a thin
shell surrounding the Earth at a height of 350\,km, 
we estimate the TEC for the line of sight from the GPS site to the
satellite using ground reception of GPS signals and knowledge of the orbital
parameters of the GPS satellites.
Having a GPS 
site near a VLBI site, we can reliably ``transfer" this estimate to the TEC
for the line of sight from the radio telescope to the radio source.
These TEC values allow us to correct the
VLBI observables for the effects of the ionosphere for any frequency.
We made such corrections for our phase-delay
data at 8.4\,GHz (Ros \et\ \cite{ros99}) for
an epoch at which there was a paucity of relevant GPS data.
The estimates of the ionospheric delays provided
by the VLBI measurements of group delay 
at 2.3 and 8.4\,GHz differ from the corresponding
delays obtained from GPS data to within
root-mean-square values
below 0.15\,ns for intercontinental baselines and 
0.10\,ns for continental ones.
Thus, we have shown, in particular, that the GPS determination of TEC 
can be successfully used in the astrometric analysis of VLBI observations.

The density of the network of GPS sites has increased dramatically 
since 1991 and the accuracy of the TEC deduced from GPS data has
improved significantly.  Given such progress, the approximations 
used in this paper are no longer necessary.
Now GPS estimates for the vertical TEC are available from virtually every
land location all of the time and thus for every VLBI observation.

The advantages of GPS compared with geostationary
beacons which used Faraday rotation to determine TECs are notable:  
global land coverage for GPS is available from 
geodetic networks (e.g., the International
GPS Geodynamics Service); 
the TEC estimates do not depend on assumptions about the Earth's
magnetic field;  L1 and L2 GPS data are available over 
the internet in standard formats (e.g., RINEX:
Receiver INdependent EXchange).
We also note that the
ionosphere cannot always be represented by a thin shell model with
good accuracy; 
moreover, more accurate models 
can be devised and suitably
parameterized given the tomography-like sampling of the ionosphere
provided by the GPS.  The biases in the GPS observables have 
to be properly corrected in the estimates of TECs.
Davies \& Hartmann (\cite{dav97}) set an upper limit of 3\,TECU for the 
present agreement of GPS TEC with the results from other methods, 
but only of few 0.01\,TECU for the relative errors of TEC estimates
over the course of some hours for any given site.
The latter represents relative changes
of delay at 8.4\,GHz of 0.2\,ps, nearly two orders of magnitude smaller than 
the delay equivalent of 
a $2\pi$ phase ambiguity.  With such accuracies and the present network 
of GPS sites, the removal of most of the effect of the ionosphere
from VLBI observations should not be difficult, although much of the 
smaller-scale ionospheric activity cannot be adequately sampled by GPS.
From dual-band difference VLBI astrometry 
and GPS data, the optical depth of the emission in
radio sources can be better
studied by comparing brightness distributions 
obtained independently at each frequency band with respect to the 
same coordinate system.
In sum, the introduction of GPS techniques should greatly improve 
the scientific results obtained from VLBI observations.

\acknowledgements  E.R.\ acknowledges a F.P.I.\ fellowship of the
Generalitat Valenciana.
We acknowledge the referee, Dr.\ R.M.\ Campbell for his very helpful
suggestions and remarks.
We are grateful to the SOPAC/IGPP, at SIO, University of California, 
San Diego (US), for kindly providing the GPS data from their GARNER archives,
and to Prof.\ R.T.\ Schilizzi for encouragement.
This work has been partially supported by the Spanish DGICYT grants 
PB\,89-0009, PB\,93-0030, and PB\,96-0782, and by the U.S.\ National
Science Foundation Grant No.\ AST 89-02087.

\end{document}